\documentclass[preprint,prl,twocolumn,showpacs,tightenlines,10pt,groupedaddress]{revtex4}
\usepackage{amsfonts}
\usepackage{amsmath}
\usepackage{amssymb}
\usepackage{graphicx}

\begin{document}

\title{Photon-atomic solitons in a Bose-Einstein condensate trapped in a
soft optical lattice }
\author{Guangjiong Dong, Jiang Zhu, Weiping Zhang}
\affiliation{State Key Laboratory of Precision Spectroscopy, Department of Physics, East
China Normal University, 3663, North Zhongshan Road, Shanghai, China}
\author{Boris A. Malomed}
\affiliation{Department of Physical Electronics, School of Electrical Engineering,
Faculty of Engineering,Tel Aviv University, Ramat Aviv 69978, Israel}
\pacs{03.75.Lm; 42.65.Tg; 05.45.Yv; 31.15.Ne}

\begin{abstract}
We investigate the ground state (GS) of a collisionless Bose-Einstein
condensate (BEC) trapped in a \textit{soft} one-dimensional optical lattice
(OL), which is formed by two counterpropagating optical beams perturbed by
the BEC density profile through the local-field effect (LFE). We show that
LFE gives rise to an envelope-deformation potential, a nonlocal potential
resulting from the phase deformation, and an effective self-interaction of
the condensate. As a result, stable \textit{photon-atomic lattice solitons},
including an optical component, in the form of the deformation of the soft
OL, in a combination with a localized matter-wave component, are generated
in the blue-detuned setting, without any direct interaction between atoms.
These self-trapped modes, which realize the system's GS, are essentially
different from the gap solitons supported by the interplay of the OL
potential and collisional interactions between atoms. A transition to
tightly bound modes from loosely bound ones occurs with the increase of the
number of atoms in the BEC.
\end{abstract}

\maketitle

The profound importance of the use of optical lattices (OLs) in atomic \cite%
{adams,bloch} and molecular \cite{dong,mp} physics is well known. These
studies assume that the interaction of an atom or molecule with
counterpropagating laser beams illuminating the gas generates a periodic
lattice potential \cite{ol} (a similar method is used for inducing virtual
lattices in photorefractive crystals, which support a great variety of
patterns \cite{Moti}). In fact, there are two aspects of the interaction of
the atomic gas with optical fields \cite{back}. First, the field produces a
density perturbation in the gas through the effective dipole potential.
Second, the OL in an inhomogeneous BEC may be affected by local variations
of the density-dependent refraction index, which is called the local-field
effect (LFE). This effect is often ignored for far-off-resonance
counterpropagating optical beams in dilute atomic and molecular gases, i.e.,
the respective OL is assumed to be \textit{rigid} \cite{adams,bloch,ol,dong}%
. However\textbf{,} the deformation of the OL by the LFE correctly explains
\cite{zhu} the asymmetric diffraction (an asymmetric momentum distribution
of scattered atoms) of a Bose-Einstein condensate (BEC) on
counterpropagating beams with unequal intensities \cite{deng2}. We call such
a deformable OL a \textit{soft lattice}.

The present work aims to demonstrate that the distortion of the blue-detuned
soft OL by the LFE induces effective interactions in the collisionless BEC,
which may be accounted for by three terms: an envelope-deformation
potential, a nonlocal potential resulting from the phase deformation, and
self-interaction. Through these terms, the LFE gives rise to (bright)
\textit{photon-atomic} \textit{lattice solitons} without the \textit{s}-wave
interaction between atoms, which represent the system's ground state (GS).
The analysis presented below reveals that a transition from the extended GS
to a strongly localized one occurs with the increase of the number of atoms
in the BEC, $N$. The analysis also shows a drastic asymmetry between blue-
and red-detuned OLs---namely, stable solitons do not emerge in the latter
case.

In this connection, it is relevant to mention that the trapping of BEC in OL
potentials provides a versatile platform for the creation of lattice
solitons \cite{ls,bloch} and the investigation of underlying mathematical
structures \cite{math}. The use of matter-wave lattice solitons as a
high-density source for atomic interferometry has been studied in mean-field
and quantum regimes \cite{bj}. Using the solitons for the design of quantum
memory and switching was proposed too \cite{nj}. Lattice solitons (alias
\textit{gap solitons}) have been created in a nearly-one-dimensional (1D)
condensate of $^{87}$Rb with repulsive inter-atomic interactions \cite%
{Markus}. Note that, although the gap solitons are, generally, dynamically
stable modes, they do not represent the GS of the interacting BEC trapped in
the OL.

Nonlinear OLs, created by a spatially-periodic modulation of the Feshbach
resonance\textbf{,} have been proposed too \cite{g}. Similar nonlinear
lattices can be created in optical media by dint of the
electromagnetically-induced transparency \cite{Konotop}. Studies of solitons
in nonlinear lattices is an active topic in photonics and matter-wave optics
\cite{rmp}. However, the LFE has not yet been addressed in the studies of
matter-wave solitons in OLs. The photon-atomic solitons generated by the
LFE, which are introduced below, i.e., localized matter-waves modes coupled
to local deformations of the underlying OL, are somewhat\textbf{\ }similar
to polariton-exciton solitons in plasmonics\textbf{\ }\cite%
{plasmonics,plasmonics-review}. They represent the GS of the system and are
thus basically different from the OL-supported gap solitons in the
self-interacting BEC\textbf{.}

We consider the atomic BEC irradiated by counter-propagating optical fields $%
E_{1}$\ and $E_{2}$\ with common frequency $\omega $.\textbf{\ }If $\omega $
is far detuned from the electronic transition frequency of atoms, the
condensate wave function $\Phi \left( \mathbf{R},T\right) $ obeys the
Gross-Pitaevskii equation \cite{back},%
\begin{equation}
i\hslash \frac{\partial \Phi }{\partial T}=\left[ -\frac{\hslash ^{2}\nabla
^{2}}{2m}+\frac{\left\vert \mathbf{d\cdot }\mathbb{E}\right\vert ^{2}}{%
\hslash \Delta }\right] \Phi ,  \label{1}
\end{equation}%
where $\mathbf{d}$ is transition matrix element, $\Delta $ is the
detuning, and normalization $\int \left\vert \Phi \right\vert
^{2}\mathbf{dr=1}$ is imposed. Direct interactions between atoms are
disregarded here, to focus on interaction mechanisms induced by the
LFE. Experimentally, the s-wave scattering length can be tuned to
nearly zero by means of the Feshbach
resonance technique \cite{fat,zero}\textbf{.} The vectorial optical field $%
\mathbb{E}$ obeys the wave equation,%
\begin{equation}
\frac{n^{2}}{c^{2}}\frac{\partial ^{2}\mathbb{E}}{\partial T^{2}}-\nabla ^{2}%
\mathbb{E}=0,  \label{2e}
\end{equation}%
with $c$ the light speed in vacuum, and the refraction index of the
condensate, $n(\mathbf{R},T)=\sqrt{1-Nd^{2}/(\varepsilon _{0}\hslash \Delta
)\left\vert \Phi \left( \mathbf{R},T\right) \right\vert ^{2}}$, which
provides for the feedback of the condensate onto the propagation of the
electromagnetic field.

We focus on the 1D situation, assuming that the matter wave is confined,
with radius $w_{\perp }$, in the transverse plane with radial variable $%
R_{\perp }$, i.e., $\Phi =\left( \pi w_{\bot }^{2}\right) ^{-1/2}\tilde{\Psi}%
(X,T)\exp \left[ -R_{\bot }^{2}/(2w_{\bot }^{2})\right] $. The confinement
may be provided by the magnetic trap, or an optical one making use of a
wavelength different from the OL-building one. Due to negligible light
scattering in the transverse direction for optical fields with a transverse
width far larger than the condensate width \cite{zs}, we omit the transverse
variation of the field and assume it polarized in the $z$ direction, $%
\mathbb{E}\left( \mathbf{R},T\right) =\tilde{E}(X)\mathbf{e}_{z}e^{-i\omega
T}$, where $\tilde{E}$ is a slowly varying amplitude. Then, multiplying Eqs.
(\ref{1}) and (\ref{2e}) by $\Phi ^{2}$\ and integrating the result in the
transverse plane, we derive the coupled 1D equations,%
\begin{equation}
i\frac{\partial \Psi }{\partial t}=-\frac{\partial ^{2}}{\partial x^{2}}\Psi
\left( x,t\right) +\frac{V_{0}}{4}\left\vert E\right\vert ^{2}\Psi ,
\label{eigen}
\end{equation}%
\begin{equation}
2i\frac{n^{2}\omega _{r}}{\omega }\frac{\partial E}{\partial t}+\frac{%
\partial ^{2}}{\partial x^{2}}E+n^{2}E=0,  \label{helmholtz}
\end{equation}%
with rescaling $E\left( x,t\right) =\tilde{E}\left( X,T\right) /E_{1}$, $%
\Psi \left( x,t\right) =\tilde{\Psi}\left( X,T\right) /\sqrt{k_{L}},$ $%
x=k_{L}X$ ,\ $t=\omega _{r}T$, $V_{0}=2d^{2}I_{1}/(\hslash ^{2}\Delta
\varepsilon _{0}c\omega _{r})$, $\omega _{r}=\hslash k_{L}^{2}/(2m)$. Here\ $%
I_{1}$ and $k_{L}$ are the intensity and wavenumber of the incident field, $%
E_{1}$, and $n\approx \sqrt{1-N\alpha \left\vert \tilde{\Psi}%
(X,T)\right\vert ^{2}}$ with $\alpha \equiv d^{2}/(\pi \varepsilon
_{0}\hslash \Delta w_{\bot }^{2})$.

Since the atomic motion is nonrelativistic ($n^{2}\omega _{r}/\omega \sim
10^{-13}$), one may drop the first term in Eq. (\ref{helmholtz}), assuming
that the incident light has the form of a long pulse in the temporal
direction. Then, a solution to Eq. (\ref{helmholtz}) is a superposition of
counterpropagating waves with slowly varying amplitudes and phase shifts
\cite{ks},
\begin{equation}
E=n^{-1/2}\left\{ A^{+}\exp \left[ i\left( x+\delta \phi \right) \right]
+A^{-}\exp \left[ -i\left( x+\delta \phi \right) \right] \right\} ,
\label{2waves}
\end{equation}%
where\textbf{\ }$\delta \phi =\int_{0}^{x}\left[ \sqrt{1-k_{L}N\alpha
\left\vert \Psi \right\vert ^{2}}-1\right] dx,$\textbf{\ }and amplitudes%
\textbf{\ }$A^{+}$\textbf{\ }and\textbf{\ }$A^{-}$\textbf{\ }of the right-
and left-propagating waves obey equations\textbf{\ }$\partial A^{\pm
}/\partial x=S^{\mp }A^{\mp }$\textbf{\ }with\textbf{\ }$S^{\pm }\equiv
(2n)^{-1}\left( dn/dx\right) \exp [\pm 2i\mathbf{(}x+\delta \phi )]$. As
shown in\textbf{\ }Ref. \cite{zhu}, the approximation based on Eqs. (\ref%
{2waves}) agrees well with underlying equation (\ref{helmholtz}).

We now focus on the basic case with $I_{1}=I_{2}$, i.e., $A^{+}=\sqrt{I^{+}}%
e^{i\theta ^{+}}$and $A^{-}=\sqrt{I^{+}}e^{i\theta ^{+}-i\delta \theta }$,
with phase difference $\delta \theta $. Seeking for stationary solutions to
Eq. (\ref{eigen}) as $\Psi \left( x,t\right) =\Psi \left( x\right) e^{-i\mu
t}$, and making use of Eq. (\ref{2waves}), one obtains
\begin{equation}
\mu \Psi =-\Psi ^{\prime \prime }+\left[ V_{0}\cos ^{2}\left( x\right)
+\delta V(x)\right] \Psi ,  \label{mu}
\end{equation}%
in which the LFE-induced potential, added to the conventional lattice
potential, is $\delta V(x)=V_{0}I^{+}n^{-1}\cos ^{2}\left( x+\delta \phi
+\delta \theta /2\right) -V_{0}\cos ^{2}x.$ For typical physical parameters
of the system (which are given below), an estimate yields $k_{L}\alpha \sim
10^{-6}$, hence $\delta \phi $\ and $\delta \theta $\ are small, and the
potential may be approximated by
\begin{equation}
\delta V(x)\approx \delta V^{\mathrm{appr}}=\delta V^{\mathrm{def}%
}(x)+\delta V^{\mathrm{nonloc}}(x)+\delta V^{\mathrm{SI}}(x).  \label{A}
\end{equation}%
Here, the term induced by the deformation of the field envelope is
\begin{equation}
\delta V^{\mathrm{def}}=V_{0}\delta I^{+}\cos ^{2}x,  \label{def}
\end{equation}%
with $\delta I^{+}\equiv I^{+}-1$; the nonlocal interaction potential is
\begin{equation}
\delta V^{\mathrm{nonloc}}=-(1/2)V_{0}I^{+}\sin \left( 2x\right) \sin \left(
\delta \Theta \right) ,  \label{nonlocal}
\end{equation}%
with $\delta \Theta \equiv 2\delta \phi +\delta \theta $ being the phase
difference between the right- and left-traveling electromagnetic waves
induced by the LFE; and
\begin{equation}
\delta V^{\mathrm{SI}}=\gamma I^{+}\left( \cos ^{2}x\right) \left\vert \Psi
\right\vert ^{2}  \label{SI}
\end{equation}%
represents the effective spatially modulated self-repulsion of the
condensate, whose strength is periodically modulated in space, with
amplitude $\gamma =Nd^{4}k_{L}I_{1}/(\pi c\varepsilon _{0}^{2}\hslash
^{3}w_{\bot }^{2}w_{r}\Delta ^{2})$.

Without terms $\delta V^{\mathrm{nonloc}}$ and $\delta V^{\mathrm{SI}}$, Eq.
(\ref{mu}) features a combination of the linear and nonlinear OLs, hence
lattice solitons \cite{bloch,ls,rmp} may be expected in this setting.
However, the deformation of the OL is a part of the present setting, i.e.,
the nonlocal and deformation potentials make the situation different from
the straightforward combination of the linear and nonlinear OLs.

\begin{figure}[tbph]
\centering\includegraphics[width=\columnwidth]{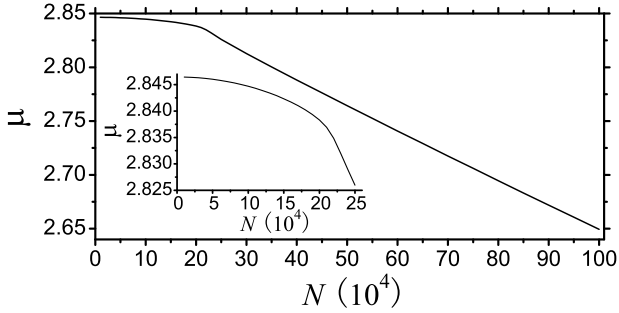}
\caption{The chemical potential of the numerically found ground
state vs. the number of atoms, $N$. The inset shows details for
relatively small values of $N$.} \label{fig2}
\end{figure}

To produce accurate results, we solved Eqs. (\ref{eigen}) and (\ref%
{helmholtz}) numerically, with parameters taken for transition 5$^{2}$S$%
_{1/2}$ $\rightarrow $ 5$^{2}$P$_{3/2}$ in $^{87}$Rb, blue detuning $\Delta
=+2$ GHz, transverse radius $10$ $\mathrm{\mu }$m, and $V_{0}=10$ (the OL
strength measured in units of the recoil energy), which corresponds to
optical intensity $I_{1}=10.26$ mW/cm$^{2}$. Chemical potential $\mu $ of
the so found GS is plotted as a function of $N$ in Fig. \ref{fig2}. When $N$
is large ($N\gtrsim 2\times 10^{5}$), $\mu $ decreases almost linearly with $%
N$, the situation being different at smaller $N$. Note that a similar linear%
\textbf{\ }dependence between $\mu $ and $N$ occurs for solitons of the 1D
nonlinear Schr\"{o}dinger equation with the nonlinearity of power $7/3$,
which occurs in some settings for interacting fermionic \cite{7/3} and
bosonic \cite{7/3-Bose} gases. In contrast, for usual gap solitons $\mu $ is
a growing, rather than decreasing, function of $N$\ \cite{ls}, although a
combination of linear and nonlinear OLs may change this \cite{HS2}.

\begin{figure}
\centering\includegraphics[width=\columnwidth]{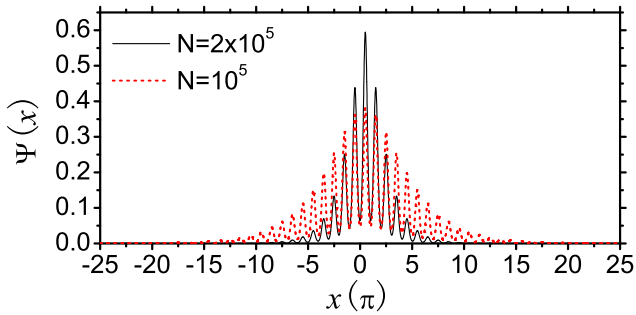}
\caption{(Color online) Wave functions of loosely bound ground
states for different numbers of atoms .}
\label{fig3}
\end{figure}

\begin{figure}
\centering\includegraphics[width=\columnwidth]{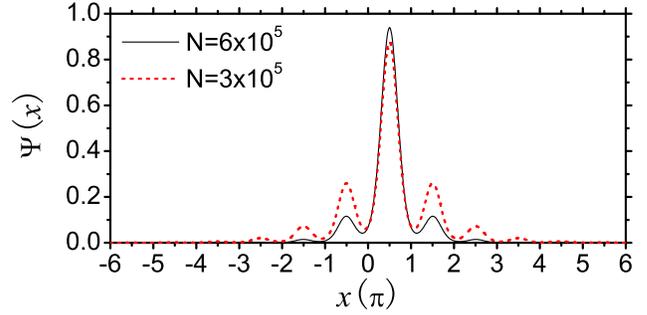}
\caption{(Color online) The ground-state wave functions for larger
$N$ than in Fig. \ref{fig3}, in the form of tightly bound solitons.}
\label{fig4}
\end{figure}

When $N$ is relatively small, the numerically found GS is loosely bound,
featuring slowly decaying spatial undulations, as shown in Fig. \ref{fig3}
for $N=10^{5}$ and $2\times 10^{5}$. In contrast, for larger $N$, such as $%
3\times 10^{5}$and $6\times 10^{5}$, which correspond to the nearly linear $%
\mu (N)$\ curve in Fig. \ref{fig2}, the GS is found in the form of stable
tightly-bound solitons, see Fig. \ref{fig4}. The range of values of $N$
considered here is realistic, as BEC can be readily created with $N$ up to $%
\sim 10^{8}$ \cite{large-N}. A closer inspection of the profiles of the
loosely and tightly bound modes demonstrates that, although undulating, they
never cross zero (unlike gap solitons in the model with the fixed OL), which
corroborates that they are GSs. The stability of the modes has been verified
by simulations of their evolution within the framework of Eqs. (\ref{eigen})
and (\ref{helmholtz}). It is relevant to note that the gradual transition
from the loosely to tightly bound GS is similar to the transformation of
solitons reported in Ref. \cite{bf}, where the effective interaction between
bosons was modified by admixing fermions to the system.

Further simulations of Eqs. (\ref{eigen}) and (\ref{helmholtz}) demonstrate
that the photon-atomic solitons are immobile (they absorb a suddenly applied
kick), which is explained by the fact that they are pinned to the underlying
OL. In that sense, they are similar to discrete solitons pinned to the
underlying lattice \cite{DNLS}.

\begin{figure}
\centering\includegraphics[width=\columnwidth]{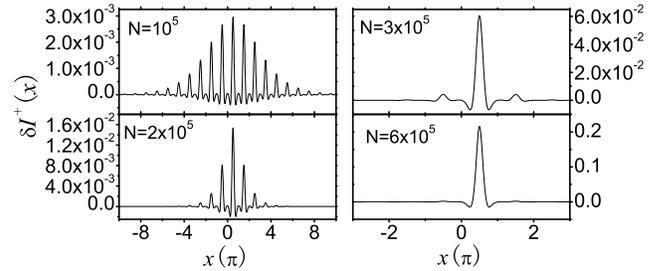}\caption{The
deformation of the envelope of the right-traveling electromagnetic
wave for different values of $N$.} \label{fig5} \end{figure}

The deformation of the envelope of the right-traveling electromagnetic wave,
$\delta I^{+}$, which is a photonic component of the GS, is plotted in Fig. %
\ref{fig5}. Local peaks of the intensity appear at $x=2n\pi +\pi /2$
(with integer $n$), coinciding with local maxima of the density in
the matter-wave component, cf. Figs. \ref{fig3} and \ref{fig4}.
Thus, the GS is indeed a coupled self-trapped photon-atomic mode (a
``symbiotic" one, cf. this concept developed for solitons in other
settings \cite{sym}).

\begin{figure}
\centering\includegraphics[width=\columnwidth]{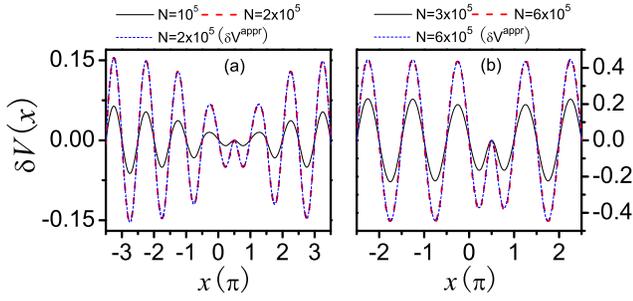}
\caption{(Color online) The numerically found LFE-induced
potential,$\protect\delta V$ for $N=10^{5}$ and $2\times 10^{5}$\
(a), and $N=$ $3\times 10^{5}$ and $6\times 10^{5}$ (b). The
approximate potential $\protect\delta V^{\mathrm{appr}}$, given by Eq. (%
\protect\ref{A}), is shown,\ for $N=2\times 10^{5}$\ and $6\times
10^{5}$, by the dotted lines.}\label{fig6}
\end{figure}

To further illustrate the intrinsic structure of the solitons, the deviation
of the effective LFE-affected potential from the periodic one induced by the
unperturbed OL, $\delta V(x)$, is plotted in Figs. \ref{fig6}(a) and \ref%
{fig6}(b). The figures feature a self-sustained \textquotedblleft valley" in
the central region, superimposed on the nearly-periodic potential.\textbf{\ }%
Note that modulated potentials, unlike strictly periodic ones, can maintain
localized states, a well-known example being the Anderson localization in
quasi-periodic potentials \cite{localization}. This analogy helps to
understand the trapping of the atomic wave function in the soliton. The
valley becomes narrower with the increase of $N$, leading to the tighter
bound modes.

\begin{figure}
\centering\includegraphics[width=\columnwidth]{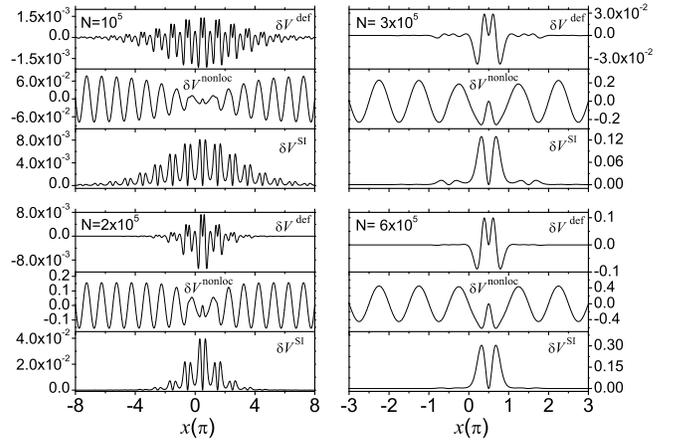}\caption{The
potential terms accounting for the envelope-deformation effect,
$\protect\delta V^{\mathrm{def}}$, the effective nonlocality,
$\protect\delta V^{\mathrm{nonloc}}$, and the light-induced
self-interaction, $\protect\delta V^{\mathrm{SI}}$, are shown for
different $N$.}\label{fig7}
\end{figure}

To check how well the full numerically found distortion of the potential, $%
\delta V(x)$, is fitted by approximation $\delta V^{\mathrm{appr}}$ given by
Eq. (\ref{A}), we plot $\delta V^{\mathrm{appr}}(x)$\ for $N=2\times 10^{5}$
and $6\times 10^{5}$ in Fig. \ref{fig6}, which shows a negligible difference
between $\delta V$ and $\delta V^{\mathrm{appr}}$. Further, the relation of
the envelope-deformation, nonlocal-interaction, and self-interaction
potential terms, $\delta V^{\mathrm{def}}$, $\delta V^{\mathrm{nonloc}}$,
and $\delta V^{\mathrm{SI}}$, in the approximate potential to the atom
number is presented in Fig. \ref{fig7}. When $N$ is small, $\delta V^{%
\mathrm{nonloc}}$\ is the dominant term. It features the above-mentioned
central \textquotedblleft valley" which supports the loosely bound GS wave
function. With the increase of $N$, the shape of the potentials simplifies,
the \textquotedblleft valley" shrinks, and terms $\delta V^{\mathrm{def}}$
and $\delta V^{\mathrm{SI}},$ featuring a trapping region, quickly grow.
Being strongly\textbf{\ }localized, they maintain tightly bound GS wave
functions, as seen in Fig. \ref{fig4}.

Additional analysis demonstrates that, quite naturally, the self-trapped
modes do not exist if direct repulsive interactions between atoms dominate
over the LFE, and, on the other hand, the usual mechanism of the formation
of 1D solitons \cite{solitons} supplants the LFE if direct attractive
interactions are strong enough. However, the Feshbach-resonance technique
makes it possible to reduce the direct interactions \cite{fat,zero}, if
necessary, and thus allow the LFE-based mechanism to manifest itself.

In the case of the red detuning of the electromagnetic waves ($\Delta <0$),
the same model does not produce solitons. In this case, the nonlinear
potential term $\delta V^{\mathrm{SI}}$\ and the linear OL potential are in
competition (recall that the sign of $\gamma \sim \Delta ^{-2}$\ is independ
on the sign of $\Delta $, unlike $V_{0}\sim \Delta ^{-1}$), which hampers
the creation of matter wave solitons. Simultaneously, a \textquotedblleft
potential barrier", rather than a trap, is generated for optical fields in
the Helmholtz equation (\ref{2e}) by the condensate refraction index $n\geq
1 $, prohibiting the creation of solitons.

In conclusion, we have studied the interaction of the collisionless BEC with
counterpropagating optical waves, including the LFE (local-field effect),
which deforms the potential of the \textquotedblleft soft" OL. This effect
leads to the deformational and nonlocal-interaction potentials, along with
the self-repulsive interaction. The so induced nonlinearity, acting along
with the linear OL, gives rise to stable photon-atomic solitons, which
realize the ground state of the system, provided that the light is
blue-detuned from the atomic transition. The increase of the number of atoms
leads to the transition from loosely bound to tightly bound solitons. It may
be interesting to extend the analysis by constructing bound complexes of
such solitons, and generalizing the analysis to the 2D setting.

These findings add to recent results demonstrating the potential of the LFE
for generating novel physical phenomena, such as non-classical matter-wave
and photonic states \cite{rit}, and \textquotedblleft photon bubbles" (which
are of great interest to astrophysics), predicted through the interaction of
diffuse light with BEC via the LFE in a magneto-optic trap \cite{bub}.
Further, the deformation of the OL by the LFE is akin to \textquotedblleft
irregular" gratings observed in asymmetric super-radiance of matter waves
for blue- and red-detuned settings \cite{ld}, while the conventional
superradiance theory \cite{th} assumes a regular OL. Thus, the theory may be
extended to explore the super-radiance of matter waves. The LFE was recently
investigated too in the framework of thermal molecular dynamics \cite{mp},
suggesting a possibility to generate thermal photon-atomic solitons in that
context.

\begin{acknowledgments}
This work was supported by the National Basic Research Program of China (973
Program) under Grants No. 2011CB921604 and 2011CB921602, the National
Natural Science Foundation of China under Grants No. 10874045, 11234003,
11034002 and 10828408, and the Program of Introducing Talents of Discipline
to Universities (B12024)\textquotedblright , as well as Research Fund for
the Doctoral Program of Higher Education of China under grant No.
20120076110010. J. Zhu acknowledges Ministry Reward for Excellent Doctors in
Academics of ECNU under Grant No. MXRZZ2011007. B.A.M. appreciates
hospitality of the Departments of Physics at the East China Normal
University and Shanghai Jiaotong University.
\end{acknowledgments}

\end{document}